\documentclass[preprint]{aastex}

\def\gapprox{\lower.4ex\hbox{$\;\buildrel >\over{\scriptstyle\sim}\;$}}
\def\lapprox{\lower.4ex\hbox{$\;\buildrel <\over{\scriptstyle\sim}\;$}}
\def\kinenergy{\varepsilon}

\slugcomment{accepted by \apj}

\shorttitle{Critical Mach Numbers in Cosmic-Ray Shocks}
\shortauthors{Becker \& Kazanas}

\begin{document}

\title{EXACT EXPRESSIONS FOR THE CRITICAL \break MACH NUMBERS
IN THE TWO-FLUID MODEL \break OF COSMIC-RAY MODIFIED SHOCKS}

\author{Peter A. Becker\altaffilmark{1}}
\affil{Center for Earth Observing and Space Research,
\break and Department of Physics and Astronomy,
\break George Mason University, Fairfax, VA 22030-4444, USA}

\and

\author{Demosthenes Kazanas\altaffilmark{2}} \affil{Laboratory for
High-Energy Astrophysics,
\break NASA Goddard Space Flight Center, Greenbelt, MD
20771, USA}

\altaffiltext{1}{pbecker@gmu.edu}
\altaffiltext{2}{kazanas@milkyway.gsfc.nasa.gov}

\begin{abstract}
The acceleration of relativistic particles due to repeated scattering
across a shock wave remains the most attractive model for the production
of energetic cosmic rays. This process has been analyzed extensively
during the past two decades using the ``two-fluid'' model of diffusive
shock acceleration. It is well known that 1, 2, or 3 distinct solutions
for the flow structure can be found depending on the upstream
parameters. Interestingly, in certain cases both smooth and
discontinuous transitions exist for the same values of the upstream
parameters. However, despite the fact that such multiple solutions to
the shock structure were known to exist, the precise nature of the
critical conditions delineating the number and character of shock
transitions has remained unclear, mainly due to the inappropriate choice
of parameters used in the determination of the upstream boundary
conditions. In this paper we derive the exact critical conditions by
reformulating the upstream boundary conditions in terms of two
individual Mach numbers defined with respect to the cosmic-ray and gas
sound speeds, respectively. The gas and cosmic-ray adiabatic indices are
assumed to remain constant throughout the flow, although they may have
arbitrary, independent values. Our results provide for the first time a
complete, analytical classification of the parameter space of shock
transitions in the two-fluid model. We use our formalism to analyze the
possible shock structures for various values of the cosmic-ray and gas
adiabatic indices. When multiple solutions are possible, we propose
using the associated entropy distributions as a means for indentifying
the most stable configuration.

\end{abstract}



\section{INTRODUCTION}

It is now widely accepted that acceleration in supernova-driven shock
waves plays an important role in the production of the observed cosmic
ray spectrum up to energies of $\sim 10^{15}\,$eV (Heavens 1984a; Ko
1995a), and it is plausible that acceleration in shocks near
stellar-size compact objects can produce most of the cosmic radiation
observed at higher energies (Jones \& Ellison 1991). In the generic
shock acceleration model, cosmic rays scatter elastically with magnetic
irregularities (MHD waves) that are frozen into the background (thermal)
gas (Gleeson \& Axford 1967; Skilling 1975). In crossing the shock,
these waves experience the same compression and deceleration as the
background gas, if the speed of the waves with respect to the
gas (roughly the Alfv\'en speed) is negligible compared with the flow
velocity (Achterberg 1987). The convergence of the scattering centers in
the shock creates a situation where the cosmic rays gain energy
systematically each time they cross the shock. Since the cosmic rays
are able to diffuse spatially, they can cross the shock many times. In
this process, an exponentially small fraction of the cosmic rays
experience an exponentially large increase in their momentum due to
repeated shock crossings. The characteristic spectrum resulting from
this first-order Fermi process is a power-law in momentum (Krymskii
1977; Bell 1978a, b; Blandford \& Ostriker 1978).

It was recognized early on that if the cosmic rays in the downstream
region carry away a significant fraction of the momentum flux supplied
by the incident (upstream) gas, then the dynamical effect of the
cosmic-ray pressure must be included in order to obtain an accurate
description of the shock structure (Axford, Leer, \& Skadron 1977). In
this scenario, the coupled nonlinear problem of the gas dynamics and the
energization of the cosmic rays must be treated in a self-consistent
manner. A great deal of attention has been focused on the ``two-fluid''
model for diffusive shock acceleration as a possible description for the
self-consistent cosmic-ray modified shock problem. In this steady-state
theory, first analyzed in detail by Drury \& V\"olk (1981, hereafter
DV), the cosmic rays and the background gas are modeled as interacting
fluids with constant specific heat ratios $\gamma_c$ and $\gamma_g$,
respectively. The coupling between the cosmic rays and the gas is
provided by MHD waves, which serve as scattering centers but are
otherwise ignored. The cosmic rays are treated as massless particles,
and second-order Fermi acceleration due to the stochastic wave
propagation is ignored. Within the context of the two-fluid model, DV
were able to demonstrate the existence of multiple (up to 3) distinct
dynamical solutions for certain upstream boundary conditions. The
solutions include flows that are smooth everywhere as well as flows that
contain discontinuous, gas-mediated ``subshocks.'' Subsequently,
multiple solutions have also been obtained in modified two-fluid models
that incorporate a source term representing the injection of seed cosmic
rays (Ko, Chan, \& Webb 1997; Zank, Webb, \& Donohue 1993). Only one
solution can be realized in a given flow, but without incorporating
additional physics one cannot determine which solution it will be.
The two-fluid model has been extended to incorporate a quantitative
treatment of the MHD wave field by McKenzie \& V\"olk (1982) and
V\"olk, Drury, \& McKenzie (1984) using a ``three-fluid'' approach.

During the intervening decades, a great deal of effort has been expended
on analyzing the structure and the stability of cosmic-ray modified
shocks (see Jones \& Ellison 1991 and Ko 1995b for reviews). Much of this
work has focused on the time-dependent behavior of the two-fluid model,
which is known to be unstable to the development of acoustic waves
(Drury \& Falle 1986; Kang, Jones, \& Ryu 1992) and magnetosonic waves
(Zank, Axford, \& McKenzie 1990). Ryu, Kang, \& Jones (1993) extended
the analysis of acoustic modes to include a secondary, Rayleigh-Taylor
instability. In most cases, it is found that the cosmic-ray pressure
distribution is not substantially modified by the instabilities.

The two-fluid theory of DV suffers from a ``closure problem'' in the
sense that there is not enough information to compute the adiabatic
indices $\gamma_g$ and $\gamma_c$ self-consistently, and therefore they
must be treated as free parameters (Achterberg, Blandford, \& Periwal
1984; Duffy, Drury, \& V\"olk 1994). This has motivated the subsequent
development of more complex theories that utilize a diffusion-convection
transport equation to solve for the cosmic-ray momentum distribution
along with the flow structure self-consistently. In these models,
``seed'' cosmic rays are either advected into the shock region from far
upstream, or injected into the gas within the shock itself.
Interestingly, Kang \& Jones (1990), Achterberg, Blandford, \& Periwal
(1984), and Malkov (1997a; 1997b) found that diffusion-convection
theories can also yield multiple dynamical solutions for certain values
of the upstream parameters, in general agreement with the two-fluid
model. Frank, Jones, \& Ryu (1995) have obtained numerical solutions to
the time-dependent diffusion-convection equation for oblique cosmic-ray
modified shocks that are in agreement with the predictions of the
steady-state two-fluid model. These studies suggest that, despite its
shortcomings, the two-fluid theory remains one of the most useful tools
available for analyzing the acceleration of cosmic rays in shocks waves
(Ko 1995b).

In their approach to modeling the diffusive acceleration of cosmic rays,
DV stated the upstream boundary conditions for the incident flow in
terms of the total Mach number,
$$
M \equiv  {v \over a} \,, \ \ \ \ \ \ 
a \equiv \sqrt{{\gamma_g \, P_g \over \rho} + {\gamma_c \, P_c \over \rho}}
\,, \eqno(1.1)
$$
and the ratio of the cosmic-ray pressure to the total pressure,
$$
Q \equiv {P_c \over P}\,, \ \ \ \ \ \ 
P \equiv P_c + P_g \,, \eqno(1.2)
$$
where $v$, $a$, $\rho$, $P$, $P_c$, and $P_g$ denote the flow velocity
of the background gas, the total sound speed, the gas density, and the
total, cosmic-ray, and gas pressures, respectively. We will use the
subscripts ``0'' and ``1'' to denote quantities associated with the
far upstream and downstream regions, respectively. DV described the
incident flow conditions by selecting values for $M_0$ and $Q_0$.
Once these parameters have been specified, the
determination of the flow structure (and in particular the number of
possible solutions) in the simplest form of the two-fluid model
requires several stages of root finding. Since the analysis
is inherently numerical in nature, the results are usually stated
only for specific upstream conditions.

The characterization of the upstream conditions in terms of $M_0$ and
$Q_0$ employed by DV turns out to be an inconvenient choice from the
point of view of finding exact critical relations describing the number
of possible flow solutions for given upstream conditions. As an
alternative approach, it is possible to work in terms of the individual
gas and cosmic-ray Mach numbers, defined respectively by
$$
M_g \equiv {v \over a_g} \,, \ \ \ \ \ \ 
M_c \equiv {v \over a_c} \,, \eqno(1.3)
$$
where
$$
a_g \equiv \sqrt{\gamma_g \, P_g \over \rho} \,, \ \ \ \ \ \ 
a_c \equiv \sqrt{\gamma_c \, P_c \over \rho} \,, \eqno(1.4)
$$
denote the gas and cosmic-ray sound speeds, respectively. According
to equation~(1.1), $a^2 = a_g^2 + a_c^2$, and therefore the Mach
numbers $M_g$ and $M_c$ are related to the total Mach number $M$
and the pressure ratio $Q$ via
$$
M^{-2} = M_g^{-2} + M_c^{-2} \,, \ \ \ \ \ \ 
Q = {\gamma_g \, M_g^2 \over \gamma_c \, M_c^2
+ \gamma_g \, M_g^2} \,, \eqno(1.5)
$$
or, equivalently,
$$
M_g = \left(1 + {\gamma_c \over \gamma_g} \, {Q \over 1-Q}
\right)^{1/2} M \,, \ \ \ \ \ \ 
M_c = \left(1 + {\gamma_g \over \gamma_c} \, {1-Q \over Q}
\right)^{1/2} M \,. \eqno(1.6)
$$
Since these equations apply everywhere in the flow, the boundary
conditions in the two-fluid model can evidently be expressed by
selecting values for {\it any two} of the four upstream parameters
$(M_0, Q_0, M_{g0}, M_{c0})$. In their work, DV described the upstream
conditions using $(M_0,Q_0)$, whereas Ko, Chan, \& Webb (1997) and
Axford, Leer, \& McKenzie (1982) used $(M_{g0},Q_0)$. Another
alternative, which apparently has not been considered before, is to use
the parameters $(M_{g0},M_{c0})$. Although these choices are all
equivalent from a physical point of view, we demonstrate below that the
set $(M_{g0},M_{c0})$ is the most advantageous mathematically because it
allows us to derive {\it exact} constraint curves that clearly delineate
the regions of various possible behavior in the parameter space of the
two-fluid model. This approach exploits the formal symmetry between the
cosmic-ray quantities and the gas quantities as they appear in the
expressions describing the asymptotic states of the flow.

The remainder of the paper is organized as follows. In \S~2 we discuss
the transport equation for the cosmic rays and derive the associated
moment equation for the variation of the cosmic-ray energy density. In
\S~3 we employ momentum and energy conservation to obtain an exact
result for the critical upstream cosmic-ray Mach number that determines
whether smooth flow is possible for a given value of the upstream gas
Mach number. In \S~4 we derive exact critical conditions for the
existence of multiple solutions containing a discontinuous, gas-mediated
subshock. The resulting curves are plotted and analyzed for various values of the
adiabatic indices $\gamma_g$ and $\gamma_c$. In \S~5 we present specific
examples of multiple-solution flows that verify the predictions made
using our analytical critical conditions. We conclude in \S~6 with a
general discussion of our results and their significance for the theory
of diffusive cosmic-ray acceleration.

\section{GOVERNING EQUATIONS}

The two-fluid model is developed by treating the cosmic rays as a fluid
with energy density comparable to that of the background gas, but
possessing negligible mass. In this section we review the basic
equations relevant for the two-fluid model. For integrity and clarity
of presentation, we also include re-derivations of a few of
the published results concerning the overall shock structure and the
nature of the transonic flow.

\subsection{\it Lagrangian Equations}

The diffusive acceleration of energetic cosmic rays due to the convergence
of scattering centers in a one-dimensional, plane-parallel flow
is described by the transport equation (Skilling 1971; 1975)
$$
{D f \over D t} = {p \over 3}\,{\partial f \over \partial p}\,
{\partial v \over \partial x} + {\partial \over \partial x}\left(
\kappa\,{\partial f \over \partial x}\right) \,, \eqno(2.1)
$$
where $p$ is the particle momentum, $v(x,t)$ is the flow velocity of the
background gas (taken to be positive in the direction of increasing $x$),
$\kappa(x,p,t)$ is the spatial diffusion coefficient, and the operator
$$
{D \over D t} \equiv {\partial \over \partial t} + v \, {\partial
\over \partial x} \eqno(2.2)
$$
expresses the comoving (Lagrangian) time derivative in the frame of
the gas. Equation~(2.1) describes the effects of Fermi
acceleration, bulk advection, and spatial diffusion on the
direction-integrated (isotropic) cosmic-ray momentum distribution
$f(x,p,t)$, which is normalized so that the total number density
of the cosmic rays is given by
$$
n_c(x,t) = \int_0^\infty 4 \pi \, p^2 \, f \, dp \,. \eqno(2.3)
$$
Note that equation~(2.1) neglects the second-order Fermi acceleration of
the cosmic rays that occurs due to stochastic wave propagation, which is
valid provided the Alfv\'en speed $v_A = B/\sqrt{4 \pi \rho}$ is
much less than the flow velocity $v$, where $B$ is the magnetic field
strength. Furthermore, equation~(2.1) does not include a particle
collision term, and therefore it is not applicable to the background
gas, which is assumed to have a thermal distribution. The momentum,
mass, and energy conservation
equations for the gas can be written in the comoving frame as
$$
{D v \over D t} = - {1 \over \rho} \, {\partial P \over \partial x} \ ,
\ \ \ \ \ \ \ \ {D \rho \over D t} = - \rho \, {\partial v \over
\partial x} \ , \ \ \ \ \ \ \ \ {D U_g \over D t} = \gamma_g \,
{U_g \over \rho} \, {D \rho \over D t} \ , \eqno(2.4)
$$
respectively,
where $U_g = P_g/(\gamma_g-1)$ is the internal energy density of
the gas. The expression for $D U_g / D t$ in equation~(2.4)
implies a purely adiabatic variation of $U_g$, and therefore it
neglects any heating or cooling of the gas due to wave generation
or damping. This adiabatic equation must be replaced with the
appropriate Rankine-Hugoniot jump conditions at a discontinuous,
gas-mediated subshock, should one occur in the flow. In the case
of a relativistic subshock, the momentum conservation
equation for the gas must be modified to reflect the anisotropy of the
pressure distribution (e.g., Kirk \& Webb 1988).

\subsection{\it Cosmic-Ray Energy Equation}

The pressure $P_c$ and the energy density $U_c$ associated with
the isotropic cosmic-ray momentum distribution $f$ are given by
(Duffy, Drury, \& V\"olk 1994)
$$
P_c(x,t) = \int_0^\infty {4 \pi \over 3} \, p^3 V \, f \, dp \,,
\ \ \ \ \ \ \ \ 
U_c(x,t) = \int_0^\infty 4 \pi \, p^2 \kinenergy \, f \, dp \,,
\eqno(2.5)
$$
where
$$
\kinenergy = (\gamma - 1) \, m c^2\,, \ \ \ \ \ \ 
V = {p \over \gamma m}\,,
\ \ \ \ \ \ 
\gamma = \sqrt{{p^2 \over m^2 c^2} + 1}\,, \eqno(2.6)
$$
denote respectively the kinetic energy, the speed, and the Lorentz
factor of a cosmic ray with momentum $p$ and mass $m$.
Although the
lower bound of integration is formally taken to be $p=0$, in practice
the cosmic rays are highly relativistic particles, and therefore $f$
vanishes for $p \lapprox m c$. If the distribution has the power-law
form $f \propto p^{-q}$, then we must have $4 < q < 5$ in order to avoid
divergence in the integrals for $P_c$ and $U_c$ (Achterberg, Blandford,
\& Periwal 1984), although this restriction can be lifted if cutoffs are
imposed at high and/or low momentum (Kang \& Jones 1990).

We can obtain a conservation equation for the cosmic-ray energy density
$U_c$ by operating on the transport equation~(2.1) with $\int_0^\infty 4
\pi \, p^2 T \, dp$, yielding
$$
{D U_c \over D t} = - \gamma_c \, U_c \, {\partial v \over \partial x}
+ {\partial \over \partial x}\left(\bar\kappa \, {\partial U_c
\over \partial x}\right)
\,, \eqno(2.7)
$$
where the mean diffusion coefficient $\bar\kappa$ is defined by (Duffy,
Drury, \& V\"olk 1994)
$$
\bar\kappa(x,t) \equiv {\int_0^\infty p^2 T \kappa \, (\partial f /
\partial x) \, dp \over \int_0^\infty p^2 T \, (\partial f /
\partial x) \, dp} \,, \eqno(2.8)
$$
and the cosmic-ray adiabatic index $\gamma_c$ is defined by
(Malkov \& V\"olk 1996)
$$
\gamma_c(x,t) \equiv {4 \over 3} + {1 \over 3}{\int_0^\infty
p^2 T \, f \, \gamma^{-1} \, dp \over
\int_0^\infty p^2 T \, f \, dp} = 1 + {P_c \over U_c}\,. \eqno(2.9)
$$
Note that in deriving equation~(2.7), we have dropped an extra term that
arises via integration by parts because it must vanish in order to
obtain finite values for $P_c$ and $U_c$. The integral expression in
equation~(2.9) indicates that $\gamma_c$ must lie in the range $4/3 \le
\gamma_c \le 5/3$. It also demonstrates that $\gamma_c$ will evolve in
response to changes in the shape of the momentum distribution $f$. The
closure problem in the two-fluid model arises because $f$ is not
calculated at all, and therefore $\gamma_c$ must be imposed rather than
computed self-consistently.

\subsection{\it Eulerian Equations}

The conservation equations can be rewritten in standard Eulerian form
as
$$
{\partial \rho \over \partial t} = - {\partial J \over \partial x}
\,, \eqno(2.10)
$$
$$
{\partial \over \partial t} \left(\rho \, v \right)  =
- {\partial I \over \partial x} \,, \eqno(2.11)
$$
$$
{\partial \over \partial t}\left({1 \over 2} \, \rho \, v^2 + U_g
+ U_c \right) = - {\partial E \over \partial x}
\,, \eqno(2.12)
$$
where the fluxes of mass, momentum, and total energy are given
respectively by
$$
J \equiv \rho \, v \,, \eqno(2.13)
$$
$$
I \equiv \rho \, v^2 + P_g + P_c \,, \eqno(2.14)
$$
$$
E \equiv {1 \over 2} \, \rho \, v^3 + v \left(P_g + U_g \right)
+ v \left(P_c + U_c \right) - \bar\kappa \, {\partial U_c \over
\partial x}\,. \eqno(2.15)
$$
The momentum and energy fluxes can be expressed in dimensionless form as
$$
{\cal I} \equiv {I \over J v_0} = u + {\cal P}_g + {\cal P}_c \,,
\eqno(2.16)
$$
$$
{\cal E} \equiv {E \over J v_0^2} = {1 \over 2} \, u^2
+ {\gamma_g \over \gamma_g-1} \, u \, {\cal P}_g
+ {\gamma_c \over \gamma_c-1} \, u \, {\cal P}_c
- {\bar\kappa \over J v_0^2} \,
{\partial \over \partial x} \left(J v_0 \, {\cal P}_c
\over \gamma_c-1 \right) \,, \eqno(2.17)
$$
where $v_0$ is the asymptotic upstream flow velocity and the
dimensionless quantities $u$, ${\cal P}_g$, and ${\cal P}_c$
are defined respectively by
$$
u \equiv {v \over v_0} \,, \ \ \ \ \ \ 
{\cal P}_g \equiv {P_g \over J v_0} \,, \ \ \ \ \ \ \ 
{\cal P}_c \equiv {P_c \over J v_0} \,. \eqno(2.18)
$$
Note that the definition of $u$ implies that the incident flow has
$u_0=1$. These relations can be used to rewrite equations~(1.3) for
the gas and cosmic-ray Mach numbers as
$$
M_g^2 = {u \over \gamma_g {\cal P}_g} \,, \ \ \ \ \ \ \ 
M_c^2 = {u \over \gamma_c {\cal P}_c} \,. \eqno(2.19)
$$

\subsection{\it The Dynamical Equation}

In this paper we shall adopt the two-fluid approximation in the form
used by DV, and therefore we assume that the adiabatic indices
$\gamma_g$ and $\gamma_c$ are each constant throughout the flow. The
assumption of constant $\gamma_g$ is probably reasonable since the
background gas is expected to remain thermal and nonrelativistic at all
locations. The assumption of constant $\gamma_c$ is more problematic, since
we expect the cosmic ray distribution to evolve throughout the flow in
response to Fermi acceleration, but it is justifiable if the ``seed''
cosmic rays are already relativistic in the far upstream region. We also
assume that a steady state prevails, so that the fluxes $J$, $I$, and
$E$ are all conserved. In this case the quantities ${\cal P}_g$ and
${\cal P}_c$ express the pressures of the two species relative to the
upstream ram pressure of the gas $\rho_0 \, v_0^2 =
J \, v_0$, where $\rho_0$ is the asymptotic upstream mass density.
The Eulerian frame in which we are working is necessarily the frame
of the shock, since that is the only frame in which the flow can appear
stationary (Becker 1998). In a steady state, the adiabatic variation of
${\cal P}_g$ implied by equation~(2.4) indicates that along any smooth
section of the flow, the gas pressure can be calculated in terms of the
velocity using
$$
{\cal P}_g = {\cal P}_{g*} \left(u \over u_* \right)^{-\gamma_g}
\,, \eqno(2.20)
$$
where ${\cal P}_{g*}$ and $u_*$ denote fiducial quantities measured
at an arbitrary, fixed location within the section of interest.
According to equation~(2.19), the associated variation of the gas
Mach number along the smooth section of the flow is given by
$$
M_g^2 = M_{g*}^2 \left(u \over u_* \right)^{1+\gamma_g} \,,
\eqno(2.21)
$$
where $M_{g*}$ denotes the gas Mach number at the fiducial location.
Substituting for ${\cal P}_g$ in equation~(2.16) using equation~(2.20)
and differentiating the result with respect to $x$ yields the dynamical
equation (Achterberg 1987; Ko, Chan, \& Webb 1997)
$$
{d u \over d x} = {d {\cal P}_c / d x \over M_g^{-2}-1} \ . \eqno(2.22)
$$
Critical points occur where the numerator and denominator vanish
simultaneously. The vanishing of the denominator implies that
$M_g = 1$ at the critical point, and therefore the critical point
is also a {\it gas sonic point} (Axford, Leer, \& McKenzie 1982).
The vanishing of the numerator implies that $d {\cal P}_c / dx=0$
at the gas sonic point.

\subsection{\it Transonic Flow Structure}

We can rewrite the dimensionless momentum flux ${\cal I}$
by using equations~(2.19) to substitute for ${\cal P}_g$ and
${\cal P}_c$ in equation~(2.16), yielding
$$
{\cal I} = u \left(1 +  {M_g^{-2} \over \gamma_g} +
{M_c^{-2} \over \gamma_c} \right) \,. \eqno(2.23)
$$
Similarly, equation~(2.17) for the dimensionless energy flux
${\cal E}$ becomes
$$
{\cal E} =  u^2 \left({1 \over 2} + {M_g^{-2} \over \gamma_g-1}
+ {M_c^{-2} \over \gamma_c-1} \right) - {1 \over \gamma_c-1} \,
{\bar \kappa \over v_0} \, {d {\cal P}_c \over d x} \,. \eqno(2.24)
$$
Using equation~(2.23) to eliminate $M_c$ in equation~(2.24) yields
for the gradient of the cosmic-ray pressure
$$
g(u) \equiv {1 \over \gamma_c-1} \, {\bar \kappa \over v_0} \,
{d {\cal P}_c \over d x} = \left({1 \over 2} - \Gamma_c \right)
u^2 + \Gamma_c \, {\cal I} \, u + \left(\Gamma_g - \Gamma_c\right)
{u^2 \over \gamma_g \, M_g^2}
- {\cal E} \,, \eqno(2.25)
$$
where
$$
\Gamma_g \equiv {\gamma_g \over \gamma_g - 1} \ , \ \ \ \ \ \ \ 
\Gamma_c \equiv {\gamma_c \over \gamma_c - 1} \ . \eqno(2.26)
$$
Along any smooth section of the flow, $g$ depends only on $u$ by
virtue of equation~(2.21), which gives $M_g$ as a function of $u$.
In the two-fluid model, the flow is assumed to become gradient-free
asymptotically, so that $d{\cal P}_c/dx \to 0$ in the far upstream and
downstream regions (DV; Ko, Chan, \& Webb 1997). The function $g$ must
therefore vanish as $|x| \to \infty$, and consequently we can express
the values of ${\cal I}$ and ${\cal E}$ in terms of the upstream Mach
numbers $M_{g0}$ and $M_{c0}$ using
$$
{\cal I} = 1 +  {M_{g0}^{-2} \over \gamma_g} +
{M_{c0}^{-2} \over \gamma_c} \, , \eqno(2.27)
$$
and
$$
{\cal E} =  {1 \over 2} + {M_{g0}^{-2} \over \gamma_g-1}
+ {M_{c0}^{-2} \over \gamma_c-1} \, , \eqno(2.28)
$$
where we have also employed the boundary condition $u_0 = 1$.

The critical nature of the dynamical equation~(2.22) implies that
$g=0$ at the gas sonic point. Hence if the flow is to pass {\it
smoothly} through a gas sonic point, then $g$ must vanish at {\it
three} locations. It follows that one of the key questions concerning
the flow structure is the determination of the number of points at
which $g=0$. We can address this question by differentiating
equation~(2.25) with respect to $u$, which yields
$$
{d g \over d u} = {u \over \gamma_c - 1} \left(M^{-2} - 1 \right)
\,, \eqno(2.29)
$$
where $M=(M_g^{-2}+M_c^{-2})^{-1/2}$ is the total Mach number, and
we have used the result
$$
{d M_g^{-2} \over d u} = - {M_g^{-2} \over u} \left(1+\gamma_g
\right) \eqno(2.30)
$$
implied by equation~(2.21). For the second derivative of $g$ we obtain
$$
{d^2 g \over d u^2} = {- 1 - \gamma_c - M_g^{-2} \, (\gamma_g
- \gamma_c) \over \gamma_c - 1} \ . \eqno(2.31)
$$
Since the cosmic rays have a higher average Lorentz factor than the
thermal background gas, $\gamma_g > \gamma_c$ (cf. eq.~2.9), and
therefore $d^2 g / d u^2 < 0$, implying that $g(u)$ is concave down as
indicated in Figure~1. Hence there are exactly {\it two} roots for $u$
that yield $g=0$, one given by the upstream velocity $u = u_0 = 1$ and
the other given by the downstream velocity, denoted by $u = u_1$. We therefore
conclude that if the flow includes a gas sonic point, then the velocity
at that point must be either $u_0$ or $u_1$. Consequently the flow cannot
pass smoothly through a gas sonic point, as first pointed out by DV.

The flows envisioned here are decelerating, and therefore the
high-velocity root $u = u_0$ is associated with the incident flow. In
this case, the fact that $g(u)$ is concave-down implies that $dg/du < 0$
in the upstream region, and therefore based on equation~(2.29) we
conclude that $M > 1$ in the upstream region and $M < 1$ in the
downstream region, with $M=1$ at the peak of the curve where $dg/du=0$.
Hence the flow {\it must} contain a sonic transition with respect to the
{\it total} sound speed $a$. In this sense, the flow is a ``shock''
whether or not it contains an actual discontinuity. For the flow to
decelerate through a shock transition, the total upstream Mach number
must therefore satisfy the condition $M_0 > 1$. This constraint also
implies that the upstream flow must be supersonic with respect to both
the gas and cosmic-ray sound speeds (i.e., $M_{g0} > 1$ and $M_{c0} >
1$), since $M_0^{-2} = M_{g0}^{-2} + M_{c0}^{-2}$. Furthermore, for a
given value of $M_{g0}$, the upstream cosmic-ray Mach number $M_{c0}$
must exceed the minimum value
$$
M_{c,\rm min} \equiv \left(1 - M_{g0}^{-2} \right)^{-1/2} \ ,
\eqno(2.32)
$$
corresponding to the limit $M_0 = 1$. The requirement that $M_{g0} > 1$
forces us to conclude that if a gas sonic point exists in the flow, then
it must be identical to the gradient-free asymptotic {\it downstream}
state. Consequently the flow must either remain supersonic everywhere
with respect to the gas, or it must cross a discontinuous, gas-mediated
subshock. If $M_g > 1$ everywhere, then the flow is completely smooth
and the gas sonic point is ``virtual,'' meaning that it exists in the
parameter space, but it does not lie along the flow trajectory. In this
case, the gas pressure evolves in a purely adiabatic fashion according
to equation~(2.20), although the total entropy of the combined system
(gas plus cosmic rays) must increase as the flow crosses the shock,
despite the fact that it is smooth. In \S~3 we derive the critical value
for the upstream cosmic-ray Mach number $M_{c0}$ that determines whether
or not smooth flow is possible for a given value of the upstream gas
Mach number $M_{g0}$.

\section{CRITICAL CONDITIONS FOR SMOOTH FLOW}

The overall structure of a cosmic-ray modified shock governed by the
dynamical equation~(2.22) can display a variety of qualitatively
different behaviors, as first pointed out by DV. Depending on the
upstream parameters, up to three distinct steady-state solutions are
possible, although only one of these can be realized in a given
situation. The possibilities include globally smooth solutions as well
as solutions containing a discontinuous, gas-mediated subshock. Smooth
flow is expected when the upstream cosmic-ray pressure $P_{c0}$ is
sufficiently large since in this case cosmic ray diffusion is able to
smooth out the discontinuity. In this
section we utilize the critical nature of the dynamical equation to
derive an analytic expression for the critical condition that determines
when smooth flow is possible, as a function of the upstream (incident)
Mach numbers $M_{c0}$ and $M_{g0}$.

\subsection{\it Critical Cosmic-Ray Mach Number}

Whether or not the flow contains a discontinuous, gas-mediated subshock,
it must be smooth in the upstream region (preceding the subshock if
one exists). We can therefore apply equation (2.21) for the variation
of $M_g$ in the upstream region, where it is convenient to use the
incident parameters $u_0=1$ and $M_{g0}$ as the fiducial quantities.
The requirement that $M_g=1$ at the gas sonic point implies that the
velocity there is given by
$$
u_s \equiv M_{g0}^{-2 / (1+\gamma_g)} \,,
\eqno(3.1)
$$
which we refer to as the ``critical velocity.'' If a sonic point
exists in the flow, then $u_s$ must correspond to the downstream
root of the equation $g(u)=0$, i.e., $u_s = u_1$. Note that the
value of $u_s$ depends only on $M_{g0}$, and consequently it is
independent of $M_{c0}$.

The asymptotic states of the flow are assumed to be gradient-free, and
the critical conditions associated with the dynamical equation~(2.22)
imply that $d{\cal P}_c/dx=0$ at the gas sonic point. The constancy of
${\cal E}$ and ${\cal I}$ therefore allows us to link upstream
quantities to quantities at the gas sonic point by using
equations~(2.23), (2.24), (2.27), and (2.28) to write
$$
{\cal I} = u_s \left(1 +  {1 \over \gamma_g} +
{M_{cs}^{-2} \over \gamma_c} \right)
= 1 +  {M_{g0}^{-2} \over \gamma_g} +
{M_{c0}^{-2} \over \gamma_c} \,, \eqno(3.2)
$$
and
$$
{\cal E} =  u_s^2 \left({1 \over 2} + {1 \over \gamma_g-1}
+ {M_{cs}^{-2} \over \gamma_c-1} \right)
= {1 \over 2} + {M_{g0}^{-2} \over \gamma_g-1}
+ {M_{c0}^{-2} \over \gamma_c-1}
\,, \eqno(3.3)
$$
respectively, where $M_{cs}$ denotes the value of the cosmic-ray
Mach number at the gas sonic point. If we substitute for $u_s$ using
equation~(3.1) and eliminate $M_{cs}$ by combining equations~(3.2)
and (3.3), we can solve for $M_{c0}$ to obtain an {\it exact expression}
for the critical upstream cosmic-ray Mach number required for the
existence of a sonic point in the asymptotic downstream region,
$$
M_{c0}=M_{cA} \equiv \left\{
{\left[\left(R_0 - 1 - {1 \over \gamma_g} \right)
M_{g0}^2 + {R_0 \over \gamma_g} \right] \gamma_c
- \left[\left({R_0^2 \over 2} - {1 \over 2} - {1 \over \gamma_g-1} \right)
M_{g0}^2 + {R_0^2 \over \gamma_g - 1} \right] \left(\gamma_c - 1 \right)
\over R_0 \, (R_0 - 1) \, M_{g0}^2}
\right\}^{-1/2}
\,, \eqno(3.4)
$$
where
$$
R_0 \equiv M_{g0}^{2/(1 + \gamma_g)} \,. \eqno(3.5)
$$
Note that $M_{cA}$ is an explicit function of the upstream gas Mach number
$M_{g0}$. The interpretation is that if $M_{c0} = M_{cA}$ for a given
value of $M_{g0}$, then the flow is everywhere supersonic with respect
to the gas sound speed $a_g$ except in the far-downstream region, where
it asymptotically approaches the gas sonic point. Surprisingly, this
simple solution for $M_{cA}$ has apparently never before appeared in the
literature, probably due to the fact that the analytical form is lost
when one works in terms of the alternative parameters $M_0$ and $Q_0$
employed by DV. This can be clearly demonstrated by using the
expressions
$$
M_{g0} = \left(1 + {\gamma_c \over \gamma_g} \, {Q_0 \over 1-Q_0}
\right)^{1/2} M_0 \ , \ \ \ \ \ \ 
M_{c0} = \left(1 + {\gamma_g \over \gamma_c} \, {1-Q_0 \over Q_0}
\right)^{1/2} M_0 \ , \eqno(3.6)
$$
to substitute for $M_{g0}$ and $M_{c0}$ in equations~(3.4) and (3.5) and
then attempting to solve the resulting equation for either $Q_0$ or
$M_0$. It is easy to convince oneself that it is not possible to express
either of these quantities explicitly in terms of the other. In Figure~2
we depict the curve $M_{c0} = M_{cA}$ in the $(M_{g0},M_{c0})$ parameter
space using equations~(3.4) and (3.5) for various values of $\gamma_g$ and
$\gamma_c$.

\subsection{\it Smooth Flow Criterion}

We have determined that smooth flow into an asymptotic downstream gas
sonic point is possible if $M_{c0} = M_{cA}$. However, in order to
obtain a complete understanding of the significance of the critical
upstream cosmic-ray Mach number $M_{cA}$, we must determine the nature
of the flow when $M_{c0} \ne M_{cA}$. The resulting flow structure can
be analyzed by perturbing around the state $M_{c0} = M_{cA}$ by taking
the derivative of the asymptotic downstream velocity $u_1$ with respect
to $M_{c0}$, holding $M_{g0}$ constant. The fact that $M_{g0}$ is held
fixed implies that the critical velocity $u_s$ also remains unchanged by
virtue of equation~(3.1). Upon differentiating, we obtain
$$
\left({\partial u_1 \over \partial M_{c0}} \right)_{M_{g0}}
= - {2 \over M_{c0}^3} \left[{\gamma_c+1 \over 2} + {(\gamma_g - \gamma_c)
\, (\gamma_g - 1 + u_1^{\gamma_g} - \gamma_g \, u_1) \over
\gamma_g \, (\gamma_g-1) \, M_{g0}^2 \, (1-u_1)^2 \, u_1^{\gamma_g}}
\right]^{-1} \,, \eqno(3.7)
$$
which is always negative since the flow decelerates, and therefore $u_1
< 1$. This indicates that if $M_{c0}$ is {\it decreased} from the value
$M_{c0}=M_{cA}$ for fixed $M_{g0}$, then the downstream velocity $u_1$
{\it increases} above the critical velocity $u_s$, and therefore the
flow is everywhere supersonic with respect to the gas sound speed. In
this case there is no gas sonic point in the flow, and consequently a
globally smooth solution is possible.
Conversely, when $M_{c0} > M_{cA}$,
a gas sonic point exists in the flow, and therefore the flow
cannot be globally smooth because that would require smooth passage
through a gas sonic point, which we have proven to be impossible. In
this case, the flow {\it must} pass through a discontinuous,
gas-mediated subshock. We conclude that globally smooth flow is
possible in the region below each of the critical curves plotted
in Figure~2.

Note that in each case there is a critical value for $M_{g0}$, denoted
by $M_{gA}$, to the right of which smooth flow is {\it always} possible
for {\it any} value of $M_{c0}$. This critical value is the solution to
the equation
$$
\left[\left(R_A - 1 - {1 \over \gamma_g} \right)
M_{gA}^2 + {R_A \over \gamma_g} \right] \gamma_c
- \left[\left({R_A^2 \over 2} - {1 \over 2} - {1 \over \gamma_g-1} \right)
M_{gA}^2 + {R_A^2 \over \gamma_g - 1} \right] \left(\gamma_c - 1 \right)
=0 \,, \eqno(3.8)
$$
where
$$
R_A \equiv M_{gA}^{2/(1 + \gamma_g)} \,, \eqno(3.9)
$$
corresponding to the limit $M_{cA} \to \infty$ in equation~(3.4). We
plot $M_{gA}$ as a function of $\gamma_g$ and $\gamma_c$ in Figure~3.
When $\gamma_g=5/3$ and $\gamma_c=4/3$, we find that $M_{gA} =
12.28$, in agreement with the numerical results of DV and Heavens
(1984b).

\section{CRITICAL CONDITIONS FOR MULTIPLE SOLUTIONS}

DV discovered that two new discontinuous solutions become
available in addition to either a smooth solution or another
discontinuous solution when the upstream total Mach number $M_0$ is
sufficiently large. Subsequent authors have confirmed the existence
of multiple dynamical solutions within the context of the two-fluid
model (Achterberg, Blandford, \& Periwal 1984; Axford, Leer, \& McKenzie
1982; Kang \& Jones 1990; Ko, Chan, \& Webb 1997; Zank, Webb, \& Donohue
1993). However, most of this work utilized numerical root-finding
procedures and therefore it fails to provide much insight into the
structure of the critical conditions that determine when multiple
solutions are possible. We revisit the problem in this section by
recasting the upstream boundary conditions using the same approach
employed in \S~3. In particular, we show that when the incident flow
conditions are stated in terms of the upstream gas and cosmic-ray Mach
numbers $M_{g0}$ and $M_{c0}$, respectively, it is possible to obtain
exact, analytical formulae for the critical curves that form the border
of the region of multiple solutions in the parameter space.

\subsection{\it Post-Subshock Flow}

The existence of multiple dynamical solutions is connected with the
presence in the flow of a discontinuous subshock mediated by the
pressure of the gas. We can therefore derive critical criteria related
to the multiple-solution phenomenon by focusing on the nature of the
flow in the post-subshock region, assuming that a subshock exists in the
flow. As we demonstrate in the Appendix, the energy, momentum, and
particle fluxes for the cosmic rays and the background gas are
independently conserved as the flow crosses the subshock. This implies
that the quantities associated with the gas satisfy the usual
Rankine-Hugoniot jump conditions, as pointed out by DV. Far downstream
from the subshock, the flow must certainly relax into a gradient-free
condition if a steady state prevails. In fact, it is possible to
demonstrate that the entire post-subshock region is gradient-free, so
that $u=$constant downstream from the subshock. This has already been
shown by DV using a geometrical approach, but it can also be easily
established using the following simple mathematical argument. First we
combine the dynamical equation~(2.22) with the definition of $g(u)$
given by equation~(2.25) to obtain the alternative form
$$
{d u \over d x} = (\gamma_c-1) \, {v_0 \over \bar \kappa} \,
{g(u) \over M_g^{-2}-1} \ . \eqno(4.1)
$$
In the post-subshock gas, $M_g < 1$ and therefore $M < 1$ regardless of
the value of $M_c$, since $M^{-2} = M_g^{-2} + M_c^{-2}$. It follows
from equation~(2.29) that $d g /d u > 0$ downstream from the subshock.
Referring to Figure~4, we wish to prove that the subshock transition
must take the velocity $u$ directly to the gradient-free asymptotic
downstream root denoted by $u_1$, so that $g = 0$ on the immediate
downstream side of the subshock. To develop the proof, let us suppose
instead that $g > 0$ in the post-subshock gas, corresponding to the
post-subshock velocity $u_a > u_1$ in Figure~4. In this case,
equation~(4.1) implies that $d u / d x > 0$ in the downstream region, so
that $u$ increases along the flow direction, evolving {\it away} from
the gradient-free root $u_1$ in the post-subshock flow. Conversely, if
$g < 0$ in the post-subshock gas (corresponding to the velocity $u_b <
u_1$ in Figure~4), then $d u / d x < 0$ in the downstream region and
consequently $u$ decreases, again evolving away from the gradient-free
root $u_1$. Hence if the flow is to be stationary, then $u$ must jump
{\it directly} to the value $u_1$, and the entire post-subshock region
must therefore be gradient-free.
This conclusion is valid within the context of the ``standard''
two-fluid model studied by DV, but Zank, Webb, \& Donohue (1993)
suggest that it may be violated in models including injection.

\subsection{\it Global Flow Structure}

We can derive an expression for $g(u)$ suitable for use in the
post-subshock region by employing equation~(2.21) to eliminate $M_g$
in equation~(2.25). This yields
$$
g(u) = \left({1 \over 2} - \Gamma_c \right)
u^2 + \Gamma_c \, {\cal I} \, u + \left(\Gamma_g - \Gamma_c\right)
{u_+^{1+\gamma_g} \, u^{1-\gamma_g} \over \gamma_g \, M_{g+}^2}
- {\cal E} \,, \eqno(4.2)
$$
where we have adopted the immediate post-subshock values $u_+$ and
$M_{g+}$ as the fiducial quantities in the smooth section of the flow
downstream from the subshock. Since we have already established that the
post-subshock flow is gradient-free, we can obtain an equation satisfied
by the post-subshock velocity $u_+$ by writing
$$
g(u_+) = 0 = \left({1 \over 2} - \Gamma_c \right)
u_+^2 + \Gamma_c \, {\cal I} \, u_+ + \left(\Gamma_g - \Gamma_c\right)
{u_+^2 \over \gamma_g \, M_{g+}^2}
- {\cal E} \,. \eqno(4.3)
$$
The gradient-free nature of the post-subshock flow also trivially
implies
$$
u_1 = u_+ \,, \ \ \ \ \ \ M_{g1} = M_{g+} \,, \eqno(4.4)
$$
where $M_{g1}$ is the asymptotic downstream gas Mach number associated
with the downstream velocity $u_1$. Equation~(4.3) can be interpreted as
a relation for the immediate {\it pre}-subshock velocity $u_-$ by
utilizing the standard Rankine-Hugoniot jump conditions (Landau \&
Lifshitz 1975)
$$
{u_+ \over u_-} = {2 + (\gamma_g - 1) \, M_{g-}^2 \over (\gamma_g + 1)
\, M_{g-}^2} \ , \ \ \ \ \ \ \ 
M_{g+}^2 = {2 + (\gamma_g - 1) \, M_{g-}^2 \over 1 - \gamma_g
+ 2 \, \gamma_g \, M_{g-}^2} \ , \eqno(4.5)
$$
where the pre-subshock gas Mach number $M_{g-}$ is related to $u_-$
and $M_{g0}$ via
$$
M_{g-}^2 = M_{g0}^2 \, u_-^{1+\gamma_g} \ ,
\eqno(4.6)
$$
which follows from equation~(2.21).

By using equations~(4.5) and (4.6) to eliminate $u_+$, $M_{g+}$, and
$M_{g-}$, we can transform equation~(4.3) into a new equation for the
pre-subshock velocity $u_-$, which we write symbolically as
$$
h(u_-) = 0 \ , \eqno(4.7)
$$
where
$$
h(u_-) \equiv \biggl[\left({1\over 2} \,
{\gamma_g + 1 \over \gamma_g - 1} - \Gamma_r -
{\Gamma_r \over \gamma_g} \, M_{g0}^{-2}
\, u_-^{-1-\gamma_g}\right) \, u_-^2
\phantom{LOTSOFSTUFF}
$$
$$
\phantom{LOTSOFSTUFF}
+ \Gamma_r \, {\cal I} \,
u_- \biggr] \, \left(\gamma_g - 1 + 2 \, M_{g0}^{-2} \, u_-^{-1-\gamma_g}
\over \gamma_g + 1 \right) - {\cal E} \ . \eqno(4.8)
$$
Recall that the constants ${\cal I}$ and ${\cal E}$ are functions of
$M_{c0}$ and $M_{g0}$ by virtue of equations~(2.27) and (2.28). In
addition to satisfying the condition $h=0$, acceptable roots for the
pre-subshock velocity $u_-$ must also exceed the critical velocity
$u_s$. This is because the flow must be supersonic with respect to the
gas before crossing the subshock, if one exists. Equation~(4.7) allows
us to solve for the pre-subshock velocity $u_-$ as a function of the
upstream cosmic-ray and gas Mach numbers $M_{c0}$ and $M_{g0}$,
respectively. In general, $u_-$ is a multi-valued function of $M_{c0}$
and $M_{g0}$, and this results in the possibility of several distinct
subshock solutions in certain regions of the parameter space.
Fortunately, additional information is also available that can be
utilized to calculate the shape of the critical curve in
$(M_{g0},M_{c0})$ space bordering the region of multiple subshock
solutions. The nature of this information becomes clear when one
examines the topology of the function $h(u_-)$ as depicted
in Figures~5 and 6 for $\gamma_g = 5/3$ and $\gamma_c = 4/3$. We
consider a sequence of situations with $M_{g0}$ held fixed and $M_{c0}$
gradually increasing from the minimum value $M_{c,\rm min}$ given by
equation~(2.32), corresponding to the limit $M_0 = 1$. Note that since
$M_{g0}$ is held constant, the critical velocity $u_s$ also remains
constant according to equation~(3.1). Two qualitatively different
behaviors are observed depending on whether or not $M_{g0}$ exceeds
$M_{gA}$, where $M_{gA} =12.28$ is the critical upstream gas Mach number
for smooth flow calculated using equation~(3.8) with $\gamma_g = 5/3$
and $\gamma_c = 4/3$.

In Figure~5{\it a} we plot $h$ as a function of
$u_-$ and $M_{c0}$ for $M_{g0} = 8$, which yields for the critical
velocity $u_s = 0.210$. In this case the minimum upstream cosmic-ray
Mach number is $M_{c,\rm min} = 1.008$. Note that initially, for
small $M_{c0}$, there is one root for $u_-$ corresponding to the
single crossing of the line $h=0$. Since this root does not exceed
$u_s$, a subshock solution is impossible and instead the flow must be
globally smooth as discussed in \S~3. The choice $M_{g0} = 8$ satisfies
the condition $M_{g0} < M_{gA}$, and therefore as $M_{c0}$ increases,
the root for $u_-$ eventually equals $u_s$, which occurs when $M_{c0} =
M_{cA} = 4.25$. In this case the subshock is located at the asymptotic
downstream limit of the flow, and therefore it is identical to the gas
sonic point. Equations~(3.1) and (4.6) indicate that the pre-subshock
gas Mach number $M_{g-} = 1$ as expected.

As we continue to increase $M_{c0}$ beyond the critical value $M_{cA}$
in Figure~5{\it a}, the root for $u_-$ exceeds $u_s$, and therefore the
smooth solution is replaced by a subshock solution. We refer to this
solution as the ``primary'' subshock solution. Since $M_{g-} > 1$ in
this region of the parameter space, the subshock plays a significant
role in modifying the flow. If $M_{c0}$ is increased further, the
primary root for $u_-$ increases slowly, the concave-down shape changes,
and a new peak begins to emerge at large $u_-$. The peak touches the
line $h=0$ when $M_{c0} = 41.65$, and therefore at this point a new
subshock root appears with the value $u_- = 0.944$. The development
of this new root can be clearly seen in Figure~5{\it b}, where we replot
$h$ at a much smaller scale. As $M_{c0}$ continues to increase, the peak
continues to rise, and the new $u_-$ root bifurcates into two roots.
Since the two new roots for $u_-$ are larger than the primary root, the
corresponding pre-subshock gas Mach numbers are also larger, and
therefore the two new subshocks are stronger than the primary subshock.
We conclude that in this region of the $(M_{g0},M_{c0})$ parameter
space, {\it three} discontinuous subshock solutions are possible,
although only one can occur in a given situation.

Another example is considered in Figure~6{\it a}, where we plot $h$ as a
function of $u_-$ and $M_{c0}$ for $M_{g0} = 13$, which yields $u_s =
0.146$ and $M_{c, \rm min} = 1.003$. In this case, $M_{g0} > M_{gA}$,
and consequently there is {\it always} a root for $u_-$ below $u_s$,
indicating that globally smooth flow is possible for all values of
$M_{c0}$. Hence the ``primary'' subshock solution never appears in this
example. However, for sufficiently large $M_{c0}$, a peak develops in
$h$ just as in Figure~5{\it b}. This peak rises with increasing $M_{c0}$
and eventually crosses the line $h = 0$ at $u_- = 0.981$ when
$M_{c0}=116$, corresponding to the appearance of a new physically
acceptable subshock root for $u_-$. This new root bifurcates into two
roots as $M_{c0}$ is increased, which can be clearly seen in
Figure~6{\it b}, where $h$ is replotted on a much smaller scale. It
follows that in this region of $(M_{g0},M_{c0})$ space, two
discontinuous solutions are possible in addition to a single globally
smooth solution. It is apparent from Figures~5 and 6 that the onset of
multiple solutions is connected with the vanishing of $h$ coupled with
the additional, simultaneous condition
$$
\left({\partial \, h \over \ \partial \, u_-}\right)_{M_{g0},M_{c0}}
= \ 0 \ , \eqno(4.9)
$$
which supplements equation~(4.7).

\subsection{\it Critical Mach Numbers for Multiple Solutions}

Equations~(4.7) and (4.9) can be manipulated to obtain explicit
expressions for the critical upstream gas and cosmic-ray Mach numbers
corresponding to the onset of multiple subshock solutions as functions
of the pre-subshock velocity $u_-$. These critical Mach numbers are
denoted by $M_{gB}$ and $M_{cB}$, respectively. The logical procedure
for obtaining the relations is straightforward, although the algebra
required is somewhat tedious. The first step in the process is to solve
equations~(4.7) and (4.9) individually to derive two separate
expressions for $M_{cB}$. Equation~(4.7) yields
$$
M_{cB} = M_{gB} \, \left({F_1 + F_2 \, M_{gB}^2 \over
F_3 + F_4 \, M_{gB}^2 + F_5 \, M_{gB}^4}\right)^{1/2}
\,, \eqno(4.10{\rm a})
$$
where
$$
F_1 \equiv 4 \, \gamma_g \, (\gamma_g - 1) \, u_-^{\gamma_g}
\,, \eqno(4.10{\rm b})
$$
$$
F_2 \equiv 2 \, \gamma_g \, (\gamma_g - 1) \, \left[
- 1 - \gamma_g + (\gamma_g - 1 ) \, u_-
\right] \, u_-^{2 \gamma_g}
\,, \eqno(4.10{\rm c})
$$
$$
F_3 \equiv - 4 \, \gamma_c \, (\gamma_g - 1) \, (u_-^{\gamma_g} - 1)
\,, \eqno(4.10{\rm d})
$$
$$
F_4 \equiv - 2 u_-^{\gamma_g} \, \bigg[
\gamma_c \, u_- \,
(u_-^{\gamma_g} - 1) +  \gamma_g \, (\gamma_g + 1) \, (u_-^{\gamma_g}
- u_-)
\phantom{LOTS_OF_STUFF}
$$
$$
\phantom{LOTS_OF_STUFF}
+ \gamma_c \, \gamma_g^2 \, (u_- - 1) \, (u_-^{\gamma_g} - 2)
- \gamma_g \, \gamma_c \, (2 - 5 \, u_- + u_-^{\gamma_g}
+ 2 \, u_-^{1+\gamma_g})
\bigg]
\,, \eqno(4.10{\rm e})
$$
$$
F_5 \equiv \gamma_g \, (\gamma_g - 1) \, (u_- - 1) \, u_-^{2 \gamma_g}
\left[- (\gamma_g + 1) \, ( \gamma_c - 1)
+ (1 + \gamma_g - 3 \, \gamma_c + \gamma_g \, \gamma_c) \, u_-
\right]
\,. \eqno(4.10{\rm f})
$$
The solution to equation~(4.9) is given by
$$
M_{cB} = M_{gB} \, \left({G_1 + G_2 \, M_{gB}^2 \over
G_3 + G_4 \, M_{gB}^2 + G_5 \, M_{gB}^4}\right)^{1/2}
\,, \eqno(4.11{\rm a})
$$
where
$$
G_1 \equiv - 2 \gamma_g^2 \, u_-^{\gamma_g}
\,, \eqno(4.11{\rm b})
$$
$$
G_2 \equiv \gamma_g \, (\gamma_g - 1) \, u_-^{1 + 2 \gamma_g}
\,, \eqno(4.11{\rm c})
$$
$$
G_3 \equiv 2 \, \gamma_g \, \gamma_c \, (- 2 + u_-^{\gamma_g})
\,, \eqno(4.11{\rm d})
$$
$$
G_4 \equiv - u_-^{\gamma_g} \, \left[
-2 \, \gamma_g^2 \, \gamma_c
+ (\gamma_g + \gamma_c - 5 \, \gamma_g \, \gamma_c + \gamma_g^2
+ 2 \, \gamma_g^2 \, \gamma_c) \, u_-
+ \gamma_c \, (\gamma_g - 1) \, u_-^{1 + \gamma_g}
\right]
\,, \eqno(4.11{\rm e})
$$
$$
G_5 \equiv \gamma_g \, u_-^{1 + 2 \gamma_g} \, \left[
- \gamma_c \, (\gamma_g - 1)
+ (1 + \gamma_g - 3 \, \gamma_c + \gamma_g \, \gamma_c) \, u_-
\right]
\,. \eqno(4.11{\rm f})
$$

The similarity of the dependences on $M_{gB}$ in equations~(4.10) and
(4.11) for $M_{cB}$ suggests that we can derive an exact solution for
$M_{gB}$ as a function of $u_-$ by equating these two expressions. The
result obtained is
$$
M_{gB} = \left({-2 \, T_1 - 2^{1/3} \, T_5 \, T_6^{-1/3}
+ 2^{-1/3} \, T_6^{1/3} \over 3 \, T_2}\right)^{1/2} \,, \eqno(4.12{\rm a})
$$
where
$$
T_1 \equiv (\gamma_g - 1) \, u_-^{2 \gamma_g} \, \bigg\{
- \gamma_g^2 \, (1+ \gamma_g) \, (1 + \gamma_c )
+ \gamma_g \, (2 + 2 \, \gamma_g \, - 7 \, \gamma_c
+ 4 \, \gamma_g \gamma_c - \gamma_g^2 \, \gamma_c) \, u_-^2
$$
$$
+ (\gamma_g + 1) \, \left[(\gamma_g - \gamma_c) \, u_-^{\gamma_g}
+ \gamma_g + \gamma_c - 5 \, \gamma_g \, \gamma_c + \gamma_g^2 + 2 \,
\gamma_g^2 \, \gamma_c
\right] \, u_-
\bigg\}
\,, \eqno(4.12{\rm b})
$$
$$
T_2 \equiv \gamma_g \, (\gamma_g - 1) \, u_-^{1 + 3 \gamma_g}
\bigg[(\gamma_c + 1) \, (\gamma_g^2 - 1)
- 2 \, (\gamma_g + 1) \, (1 + \gamma_g - 3 \, \gamma_c
+ \gamma_g \, \gamma_c) \, u_-
$$
$$+ (\gamma_g - 1) \, (1 + \gamma_g - 3 \, \gamma_c
+ \gamma_g \, \gamma_c) \, u_-^2
\bigg]
\,, \eqno(4.12{\rm d})
$$
$$
T_3 \equiv 2 \, \gamma_c \, (1 - \gamma_g^2) \,
+ (5 \, \gamma_c - 1 - \gamma_g - 5 \, \gamma_g \, \gamma_c
+2 \, \gamma_g^2 \, \gamma_c) \, u_-
+ (\gamma_g - \gamma_c - \gamma_g \, \gamma_c + \gamma_g^2)
\, u_-^{\gamma_g} \,, \eqno(4.12{\rm c})
$$
$$
T_4 \equiv 216 \, \gamma_g \, \gamma_c \, (\gamma_g - 1) \, T_2^2
- 16 \, T_1^3 - 72 \, \gamma_g \, T_1 \, T_2 \, T_3 \, u_-^{\gamma_g}
\,, \eqno(4.12{\rm e})
$$
$$
T_5 \equiv - 4 \, T_1^2 - 12 \, \gamma_g \, T_2 \, T_3 \, u_-^{\gamma_g}
\,, \eqno(4.12{\rm f})
$$
$$
T_6 \equiv T_4 + \left(T_4^2 + 4 \, T_5^3\right)^{1/2}
\,. \eqno(4.12{\rm g})
$$
Equation~(4.12) can be used to evaluate the critical upstream gas Mach
number $M_{gB}$ as a function of the pre-subshock velocity $u_-$. Once
$M_{gB}$ is determined, we can calculate the corresponding critical
upstream cosmic-ray Mach number $M_{cB}$ using either equation~(4.10) or
equation~(4.11), which both yield the same result. Hence
equations~(4.10), (4.11), and (4.12) provide a direct means for
calculating $M_{gB}$ and $M_{cB}$ as exact functions of $u_-$. As an
example, setting $\gamma_g = 5/3$, $\gamma_c = 4/3$, and $u_- = 0.944$
yields $M_{gB} = 8$ and $M_{cB} = 41.65$, in agreement with the results
plotted in Figure~5. Likewise, setting $\gamma_g = 5/3$, $\gamma_c =
4/3$, and $u_- = 0.981$ yields $M_{gB} = 13$ and $M_{cB} = 116$, in
agreement with Figure~6.

Although our expressions for $M_{gB}$ and $M_{cB}$ simplify considerably
in the special case $\gamma_g = 5/3$, $\gamma_c = 4/3$, we have chosen
to derive results valid for general (but constant) values of $\gamma_g$
and $\gamma_c$ in order to obtain a full understanding of the
sensitivity of the critical Mach numbers to variations in the adiabatic
indices. While admittedly somewhat complex, these equations can
nonetheless be evaluated using a hand calculator, and replace the
requirement of utilizing the root-finding techniques employed in most
previous investigations of the multiple-solution phenomenon in the
two-fluid model.

In Figure~7 we plot the critical curves for the occurrence of multiple
solutions using various values of $\gamma_g$ and $\gamma_c$. This is
accomplished by evaluating $M_{gB}$ and $M_{cB}$ as parametric functions
of the pre-subshock velocity $u_-$ using equation~(4.12) along with
either equation~(4.10) or (4.11). The critical curves denote the
boundary of the wedge-shaped {\it multiple-solution region}. Inside this
region, two new subshock solutions are possible, along with either the
primary subshock solution or the globally smooth solution. Outside this
region, the flow is either described by the primary subshock solution or
else it is globally smooth. The lower-left-hand corner of the
multiple-solution region curves to the left and culminates in a sharp
cusp. The presence of the cusp implies that there is a minimum value of
$M_{g0}$, below which multiple solutions are never possible for any
value of $M_{c0}$. Note that the multiple-solution region becomes very
narrow as the values of $\gamma_g$ and $\gamma_c$ approach each other,
suggesting that the physically allowed solutions converge on a single
solution as the thermodynamic properties of the two populations of
particles (background gas and cosmic rays) become more similar to each
other. In the limit $\gamma_g = \gamma_c$, multiple solutions are not
allowed at all, and the single available solution is either smooth or
discontinuous depending on the values of $M_{g0}$ and $M_{c0}$.

In Figure~8 we plot all of the various critical curves for the
physically important case of an ultrarelativistic cosmic-ray
distribution ($\gamma_c = 4/3$) combined with a nonrelativistic
background gas ($\gamma_g = 5/3$). This is the most fully
self-consistent example of the two-fluid model, since in this case we
expect that the adiabatic indices will remain constant throughout the
flow, as is assumed in the model (see eq.~[2.9]). The area of overlap
between the multiple-solution region and the smooth-solution region
implies the existence of four distinctly different domains within the
$(M_{g0},M_{c0})$ parameter space. Within Domain~I, which lies outside
both the multiple-solution region and the smooth-solution region, the
flow must be discontinuous, with exactly one (i.e., the primary
subshock) solution available. Domain~II lies inside the
multiple-solution region and outside the smooth-solution region, and
therefore within this area of the parameter space, {\it three} distinct
subshock solutions are possible, while globally smooth flow is
impossible. Domain~III is formed by the intersection of the
multiple-solution region and the smooth-solution region, and therefore
two discontinuous subshock solutions are available as well as one
globally smooth solution. Finally, Domain~IV lies outside the
multiple-solution region and within the smooth-flow region, and
therefore one globally smooth flow solution is possible. The
existence of Domain~IV is consistent with our expectation that smooth
flow will occur for sufficiently large values of the upstream
cosmic-ray pressure $P_{c0}$ due to diffusion of the cosmic rays.

If we consider a trajectory through the $(M_{g0},M_{c0})$ parameter
space that crosses into the multiple-solution region, then the sequence
of appearance of the subshock roots for $u_-$ depends on whether the
boundary is crossed through the ``top'' or ``bottom'' arcs of the wedge.
We illustrate this phenomenon for $\gamma_g = 5/3$ and $\gamma_c = 4/3$
in Figure~8, where we consider two possible paths approaching the point
$P$ inside the multiple-solution region, starting outside at either
points $Q$ or $R$. The two paths cross the multiple-solution boundary on
different sides of the wedge. In Figure~9 we plot $h(u_-)$ along the
segment $RP$ (with $M_{g0} = 6.5$), and in Figure~10 we plot $h(u_-)$
along the segment $QP$ (with $M_{c0} = 45$). Note that the primary
subshock root for $u_-$ already exists at points $Q$ and $R$ since they
both lie inside Domain~I. When the lower section of the
multiple-solution boundary is crossed along segment $RP$ ($M_{c0} =
M_{cB} = 25.9$ in Fig.~9), two new roots for $u_-$ appear at larger
values than the primary root, which is the same sequence observed in
Figure~5. However, when the {\it upper} section of the boundary is
crossed along segment $QP$ ($M_{g0} = M_{gB} = 5.83$ in Fig.~10), the
two new roots for $u_-$ appear at {\it smaller} values than the primary
root. Hence the order of appearance of the subshock roots for $u_-$ is
different along each path. Despite this path dependence, the actual set
of roots obtained at point $P$ is the same regardless of the approach
path taken, and therefore there is no ambiguity regarding the possible
subshock solutions available at any point in the $(M_{g0},M_{c0})$
parameter space.

In Figure~11 we present summary plots that include all of the various
critical curves derived in \S\S~3 and 4 for several different values of
$\gamma_g$ and $\gamma_c$. Note that the area of overlap between the
multiple-solution region and the smooth-solution region observed when
$\gamma_g = 5/3$ and $\gamma_c = 4/3$ rapidly disappears when the
difference between $\gamma_g$ and $\gamma_c$ is reduced, due to the
increasing similarity between the thermodynamic properties of the cosmic
rays and the background gas. Ko, Chan, \& Webb (1997) and Bulanov \&
Sokolov (1984) have obtained similar parameter space plots depicting the
critical curves for smooth flow and for the onset of multiple solutions.
The parameters used to describe the incident flow conditions are
$(M_{g0},Q_0)$ in the case of Ko, Chan, \& Webb (1997) and $(M_0,Q_0)$
for Bulanov \& Sokolov (1984), where $Q_0$ and $M_0$ refer to the
incident pressure ratio and total Mach number, respectively (see
eqs.~[1.1] and [1.2]). When these parameters are employed directly
instead of the
quantities $(M_{g0},M_{c0})$ utilized in our work, the critical curves
must be determined by numerical root-finding.
However, if desired, equations~(1.5) can be used to transform
our exact solutions for the critical curves from the $(M_{g0},M_{c0})$ space
to the $(M_{g0},Q_0)$
and $(M_0,Q_0)$ spaces. This procedure was used to obtain the results depicted
in Figure~12, which are identical to those presented by Ko, Chan,
\& Webb (1997) and Bulanov \& Sokolov (1984).

\section{EXAMPLES}

Our interpretation of the various critical Mach numbers derived in \S\S~3
and 4 can be verified by performing specific calculations of the flow
dynamics for a few different values of the upstream Mach numbers
$M_{g0}$ and $M_{c0}$. We shall focus here on cases with $\gamma_g =
5/3$ and $\gamma_c = 4/3$, but this is not essential. Along smooth
sections of the flow, we can calculate the velocity profile by
integrating the dynamical equation obtained by combining
equations~(2.21) and (4.1), which yields
$$
{d u \over d \tau} = (\gamma_c - 1) \, g(u) \, \left[M_{g*}^{-2} \,
\left( u \over u_*
\right)^{-1-\gamma_g} - 1 \right]^{-1} \ , \eqno(5.1)
$$
where
$$
g(u) = \left({1 \over 2} - \Gamma_c \right)
u^2 + \Gamma_c \, {\cal I} \, u + \left(\Gamma_g - \Gamma_c\right)
{u_*^{1+\gamma_g} \, u^{1-\gamma_g} \over \gamma_g \, M_{g*}^2}
- {\cal E} \,, \eqno(5.2)
$$
and we have introduced the new spatial variable
$$
\tau(x) \equiv v_0 \int_0^x \, {dx' \over
\bar \kappa(x')} \ . \eqno(5.3)
$$
The quantities $M_{g*}$ and $u_*$ are fiducial parameters measured
at an arbitrary location along the smooth section of interest, and
the constants ${\cal I}$ and ${\cal E}$ are functions of the incident
Mach numbers $M_{g0}$ and $M_{c0}$ via equations~(2.27) and (2.28).
As discussed in \S~3, if $M_{c0} < M_{cA}$, then the flow is {\it
everywhere} supersonic with respect to the gas, and therefore we can
use equation~(5.1) to describe the structure of the {\it entire} flow
by setting the fiducial parameters $M_{g*}$ and $u_*$ equal to the
asymptotic upstream quantities $M_{g0}$ and $u_0=1$, respectively.
Conversely, if $M_{c0} > M_{cA}$, then the flow must contain a
discontinuous, gas-mediated subshock. In this case the upstream
region is governed by equation~(5.1) with $M_{g*} = M_{g0}$ and
$u_* = u_0 = 1$, and the downstream region is governed by equation~(5.1)
with $M_{g*} = M_{g+}$ and $u_* = u_+$. The pre-subshock and
post-subshock regions are linked by the jump conditions given
by equations~(4.5), which determine $M_{g+}$ and $u_+$ and yield
$g=0$ in the entire downstream region as expressed by equation~(4.3).
In order to illustrate the various possible behaviors predicted
by our critical conditions, we shall present one example from each of
the four domains discussed in \S~IV (see Fig.~8) calculated using
$\gamma_g = 5/3$ and $\gamma_c = 4/3$.

In Figure~13 we plot the velocity profile $u(\tau)$
obtained by integrating the dynamical equation~(5.1) with $M_{g0} = 4$
and $M_{c0} = 4$. According to Figure~8, this point lies within Domain~I,
and therefore we expect to find a single subshock solution, and no
globally smooth solution. Equation~(2.25) confirms that in this case
the downstream root for $g(u)$ is $u = 0.27$, and therefore smooth
flow is impossible since this does not exceed the critical velocity
$u_s = 0.35$. Analysis of equation~(4.7) yields a single acceptable
value for the pre-subshock velocity root $u_- = 0.50$, with an
associated pre-subshock gas Mach number $M_{g-} = 1.60$. Figure~13
also includes plots of the gas Mach number $M_g(\tau)$ obtained using
equation~(2.21), the dimensionless gas pressure ${\cal P}_g(\tau)$
obtained using equation~(2.20), and the dimensionless cosmic-ray pressure
${\cal P}_c(\tau)$ obtained using equation~(2.16). The first three
quantities exhibit a jump at the subshock, whereas the cosmic-ray
pressure is continuous since a jump in this quantity would imply an
infinite energy flux. The overall compression ratio is $1/u_1 = 3.66 \,,
$ and the pressures increase by the factors ${\cal P}_{g1}/{\cal P}_{g0}
= 9.27$ and ${\cal P}_{c1}/ {\cal P}_{c0} = 9.89$ between the asymptotic
upstream and downstream regions. Recall that ${\cal P}_g$ and ${\cal
P}_c$ express the pressures of the two species divided by the upstream
ram pressure of the gas. Hence in this example the two species each
absorb comparable fractions of the upstream ram pressure. Note that the
increase in the cosmic-ray pressure is entirely due to the smooth part
of the transition upstream from the discontinuous subshock, while the
gas pressure experiences most of its increase in crossing the subshock.

In Figure~14 we set $M_{g0} = 6$ and $M_{c0} = 60$, corresponding to
Domain~II in Figure~8. We therefore expect to find three distinct,
physically acceptable subshock solutions associated with these parameter
values. Analysis of equation~(2.25) verifies that no smooth solutions
are possible in this case because the downstream root $u = 0.18$ is less
than the critical velocity $u_s = 0.26$. Equation~(4.7) indicates that
the three acceptable roots for the pre-subshock velocity are $u_- = 0.53
\,, 0.72 \,, 0.99 \,,$ with associated pre-subshock gas Mach numbers
$M_{g-} = 2.54 \,, 3.84 \,, 5.94 \,,$ respectively. The gas and cosmic-ray
pressures increase by the factors ${\cal P}_{g1}/{\cal P}_{g0} = 23 \,,
32 \,, 44 \,,$ and ${\cal P}_{c1}/{\cal P}_{c0} = 2126 \,, 1306 \,, 37 \,,$
and the overall compression ratios are $1/u_1 = 5.20 \,, 4.64 \,, 3.72$.
Note that the solution with the {\it largest} cosmic-ray pressure has the
{\it largest} overall compression ratio and the {\it weakest} subshock.
This is due to the fact that the cosmic-ray pressure is amplified in the
smooth-flow region upstream from the subshock, and this region is most
extended when the flow does not encounter a subshock until the smallest
possible value of $M_{g-}$. Conversely, the solution with the largest
gas pressure is the one with the strongest subshock and the smallest
overall compression ratio.

In Figure~15 we use the upstream parameters $M_{g0} = 12.5$ and $M_{c0}
= 200$. This point lies within Domain~III in Figure~8, and therefore we
expect to find two distinct subshock solutions along with one globally
smooth solution. Equation~(2.25) confirms that in this case a smooth
solution is possible since the downstream root $u = 0.152$ exceeds the
critical velocity $u_s = 0.150$. Analysis of equation~(4.7) yields two
acceptable values for the pre-subshock velocity given by $u_- = 0.962
\,, 0.997$. The corresponding pre-subshock gas Mach numbers are $M_{g-}
= 11.9 \,, 12.4 \,,$ respectively. In the discontinuous (subshock)
solutions the gas and cosmic-ray pressures increase by the factors
${\cal P}_{g1} /{\cal P}_{g0} = 188 \,, 194 \,,$ and ${\cal P}_{c1}/
{\cal P}_{c0} = 2011 \,, 170 \,, $ and the overall compression ratios
are $1/u_1 = 4.07 \,, 3.94 \,$. The subshocks in this example are
strong, and therefore in the solutions containing a subshock most of the
deceleration is due to the buildup of the pressure of the background
gas, rather than the cosmic-ray pressure. Hence both of the subshock
solutions are gas-dominated. In the globally smooth solution, which
is cosmic-ray dominated, the pressures increase by the factors ${\cal
P}_{g1}/ {\cal P}_{g0} = 23$ and ${\cal P}_{c1}/{\cal P}_{c0} = 40702$,
and the compression ratio is $1/u_1 = 6.57$. In this case the cosmic
rays absorb almost all of the ram pressure of the upstream gas.

Finally, in Figure~16 we set $M_{g0} = 14$ and $M_{c0} = 20$. According
to Figure~8, this point lies within Domain~IV, and therefore we expect
that only a single, globally-smooth solution exists for these upstream
parameters. This prediction is verified by equation~(4.7), which
confirms that no acceptable subshock roots for $u_-$ exists.
Furthermore, equation~(2.25) indicates that smooth flow is possible
since the downstream root $u = 0.15$ exceeds the critical velocity $u_s
= 0.14$. Hence the only acceptable solution is globally smooth, with the
pressure increases given by ${\cal P}_{g1}/{\cal P}_{g0} = 23$
and ${\cal P}_{c1}/{\cal P}_{c0} = 417$. The overall compression
ratio is $1/u_1 = 6.56$ and the flow is cosmic-ray dominated.

\section{DISCUSSION}

In this paper we have obtained a number of new analytical results
describing the critical behavior of the two-fluid model for cosmic-ray
modified shocks. It is well known that in this model, up to three
distinct solutions are possible for a given set of upstream boundary
conditions. The behaviors of the various solutions can be quite diverse,
including flows that are smooth everywhere as well as flows that contain
a discontinuous, gas-mediated subshock. The traditional approach to the
problem of determining the types of possible solutions, employed by Ko,
Chan, \& Webb (1997) and Bulanov \& Sokolov (1984), is based on stating
the upstream boundary conditions in terms of the incident total Mach
number $M_0$ and the incident pressure ratio $Q_0$ (see eqs.~[1.1] and
[1.2]). In this approach the determination of the available solution
types requires several steps of root-finding, and there is no
possibility of obtaining analytical expressions for the critical
relationships.

The analysis presented here utilizes a fresh approach based
upon a new parameterization of the boundary conditions in terms of the
upstream gas and cosmic-ray Mach numbers $M_{g0}$ and $M_{c0}$,
respectively. The analytical results we have obtained in \S\S~3 and 4
for the critical upstream Mach numbers expressed by equations~(3.4),
(4.11), and (4.12) provide for the first time a systematic classification
of the entire parameter space for the two-fluid model, which remains one
of the most powerful and practical means available for studying the problem
of cosmic-ray modified shocks. These expressions eliminate
the need for complex root-finding procedures in order to understand the
possible flow dynamics for a given set of upstream boundary conditions,
and are made possible by the symmetry between the gas and cosmic-ray
parameters as they appear in the expressions describing the asymptotic
upstream and downstream states of the flow. We have compared our
quantitative results with those of Ko, Chan, \& Webb (1997) and
Bulanov \& Sokolov (1984), and they are found to be consistent.
The results are valid for arbitrary (but constant) values of the
gas and cosmic-ray adiabatic indices $\gamma_g$ and $\gamma_c$,
respectively. In \S~5 we have presented numerical examples of flow
structures obtained in each of the four parameter space domains defined
in Figure~8. These examples verify the predictions made using our new
expressions for the critical Mach numbers, and confirm that the largest
overall compression ratios are obtained in the globally-smooth, cosmic-ray
dominated cases.

The existence of multiple distinct solutions for a single set of
upstream boundary conditions demands that we include additional physics
in order to determine which solution is actually realized in a given
situation. This question has been addressed by numerous authors using
various forms of stability analysis as well as fully time-dependent
calculations. DV speculated that when three distinct solutions are
allowed (in Domains II and III of the parameter space plotted in
Fig.~8), the solution with the intermediate value of the cosmic-ray
pressure ${\cal P}_c$ will be unstable. The argument is based on the idea
that if the cosmic-ray pressure were to increase slightly due to a small
perturbation, then the gas would suffer additional deceleration, leading
to a further increase in the cosmic-ray pressure. This nonlinear process
would drive the flow towards the steady-state solution with the largest
value of ${\cal P}_c$. Conversely, a small decrease in the cosmic-ray
pressure would decrease the deceleration, leading to a smaller value for
the cosmic-ray pressure. In this case the flow would be driven towards
the steady-state solution with the smallest value of ${\cal P}_c$.
Recently, Mond \& Drury (1998) have suggested that this type of behavior
may be realized as a consequence of a corrugational instability.
Other authors (e.g., Drury \& Falle 1986; Kang, Jones, \& Ryu 1992;
Zank, Axford, \& McKenzie 1990; Ryu, Kang, \& Jones 1993) have argued
that the globally smooth, cosmic-ray dominated solutions are unstable
to the evolution of MHD waves in certain situations. Jones \&
Ellison (1991) suggest that even when formally stable, the smooth solutions
may not be realizable in nature.
On the other hand, Donohue, Zank, \& Webb (1994) report time-dependent
simulations which seem to indicate that
the smooth, cosmic-ray dominated solution is indeed the
preferred steady-state solution in certain regions of the parameter space.
Hence, despite the fact that much effort has been expended in analyzing the
stability properties of cosmic-ray modified shocks, there is still no
clear consensus regarding which of the steady-state solutions (if any)
is stable and therefore physically observable for an arbitrary set of
upstream conditions.

In light of the rather contradictory state of affairs regarding the
stability properties of the various possible dynamical solutions,
we propose a new form of {\it
entropy-based} stability analysis. In this method, the entropy of the
cosmic-ray distribution is calculated by first solving the transport
equation~(2.1) for the cosmic-ray distribution $f$ and then integrating
to obtain the Boltzmann entropy per cosmic ray,
$$
\Sigma_c \equiv -k \int_0^\infty 4 \pi \, p^2 \, F \, \ln F \, dp
- k \ln \hbar^3 + k \ln V \ ,
\eqno(6.1)
$$
where $k$ is Boltzmann's constant, $n_c$ is the cosmic-ray number
density, $\hbar$ is Planck's constant, $V$ is the system volume,
and $F \equiv f/n_c$. According to equation~(2.3),
$4 \pi \, p^2 \, F(p,x) \, dp$ gives the probability that a
randomly selected cosmic ray at location $x$ has momentum in the
range between $p$ and $p + dp$. The cosmic-ray entropy
density $S_c$ is computed using
$$
S_c = n_c \, \Sigma_c - k n_c \, \ln \left(n_c V\right)
+ k n_c \ , \eqno(6.2)
$$
where the final two terms stem from the fundamental indistinguishability of
the cosmic ray particles (of like composition), and is necessary in
order to avoid the Gibbs paradox (Reif 1965). The inconvenient
reference to the system volume $V$ can be removed by combining
equations~(6.1) and (6.2) to obtain the equivalent expression
$$
S_c \equiv -k n_c \int_0^\infty 4 \pi \, p^2 \, F \, \ln F \, d p
- k n_c \ln\left(n_c \hbar^3\right) + k n_c \ .
\eqno(6.3)
$$
The total entropy per particle $\Sigma_{\rm tot}$ for the combined gas-particle
system is calculated using
$$
\Sigma_{\rm tot} = {S_c + S_g \over n_c + n_g} \ , \eqno(6.4)
$$
where $S_g$ and $n_g$ denote the entropy density and the number density
of the background gas, respectively. One may reasonably hypothesize
that the state with the largest value for $\Sigma_{\rm tot}$ will be the
preferred state in nature. This criterion may prove
useful for identifying the most stable solution when multiple solutions
are available. We plan to pursue this line of inquiry in future work.
The results may shed new light on the structure of cosmic-ray modified
shocks.

The authors acknowledge several stimulating conversations with Frank
Jones and Truong Le. The authors are also grateful to the anonymous
referee for several useful comments.
PAB also acknowledges support and hospitality
from NASA via the NASA/ASEE Summer Faculty Fellowship program at Goddard
Space Flight Center.

\vfil
\eject

\centerline{\bf APPENDIX}
\bigskip

In this section we demonstrate that the cosmic-ray energy flux must be
continuous across a velocity discontinuity (subshock), if one is present
in the flow. This is turn implies that the subshock must be mediated
entirely by the pressure of the background gas, and therefore the
discontinuity is governed by the classical Rankine-Hugoniot jump
conditions. The cosmic-ray energy flux in the $\hat x$ direction is
given by
$$
F_c = -\bar\kappa \, {d U_c \over d x} + \gamma_c \, v \, U_c
\ . \eqno(\rm A1)
$$
In a steady-state, the cosmic-ray energy equation~(2.7) reduces to
$$
v \, {d U_c \over d x} = - \gamma_c \, U_c \, {d v \over d x}
+ {d \over d x}\left(\bar\kappa \, {d U_c \over d x}\right)
\ , \eqno(\rm A2)
$$
which can be combined with equation~(A1) to express the derivative
of $F_c$ as
$$
{d F_c \over d x} = (\gamma_c - 1) \, v \, {d U_c \over d x}
\ . \eqno(\rm A3)
$$
Integration in the vicinity of the subshock located at $x=x_0$ yields
$$
\lim_{\epsilon \to 0} \, \int_{x_0 - \epsilon}^{x_0 + \epsilon}
{d F_c \over d x} \, dx =
\lim_{\epsilon \to 0} \, \int_{x_0 - \epsilon}^{x_0 + \epsilon}
(\gamma_c - 1) \, v \, {d U_c \over d x} \, dx \ . \eqno(\rm A4)
$$
In order to avoid unphysical divergence of $F_c$ at the subshock,
$U_c$ must be continuous, and therefore the integrand on the
right-hand side of equation~(A4) is no more singular than a
step function. This implies that in the limit $\epsilon \to 0$, the
right-hand side vanishes, leaving
$$
\Delta F_c \equiv \lim_{\epsilon \to 0}
\ F_c(x_0 + \epsilon) - F_c(x_0 - \epsilon) = 0 \ . \eqno(\rm A5)
$$
Hence $F_c$ remains constant across the subshock. The energy,
momentum, and particle fluxes of the gas and the cosmic rays are
therefore independently conserved across the discontinuity. This allows
us to use the standard Rankine-Hugoniot jump conditions to describe
the subshock transition in equations~(4.5).

\eject

\centerline{\bf FIGURE CAPTIONS}
\bigskip

Fig. 1. -- Topology of the function $g(u)$ given by eq.~[2.25].
The quantities $u_0$ and $u_1$ respectively denote the upstream
and downstream roots for the velocity in a globally smooth solution.

\bigskip

Fig. 2. -- Critical upstream cosmic-ray Mach number $M_{cA}$
(eq.~[3.4]) for smooth flow plotted as a function of the upstream gas Mach
number $M_{g0}$ in the $(M_{g0},M_{c0})$ parameter space
for ({\it a}) $\gamma_g = 5/3$ and various values of
$\gamma_c$ as indicated; ({\it b}) $\gamma_c = 4/3$ and various values
of $\gamma_g$ as indicated. Smooth flow is not possible in the
region above each curve.

\bigskip

Fig. 3. -- Critical upstream gas Mach number for smooth flow
$M_{gA}$ (eq.~[3.8]) plotted as a function of the gas and
cosmic-ray adiabatic indices $\gamma_g$ and $\gamma_c$, respectively.
When $M_{g0} > M_{gA}$, smooth flow is possible for any value
of $M_{c0}$.

\bigskip

Fig. 4. -- Schematic depiction of the function $g(u)$ (eq.~[2.25]).
If the flow contains a discontinuous, gas-mediated subshock, then
the velocity must jump {\it directly} to the final asymptotic value $u_1$
in crossing the shock. Otherwise the flow is unstable (see the
discussion in the text).

\bigskip

Fig. 5. -- Function $h(u_-)$ (eq.~[4.8]) is plotted for the parameters
({\it a}) $M_{g0} = 8$, $\gamma_g = 5/3$, $\gamma_c = 4/3$. The
value of $M_{c0}$ is indicated for each curve.
In this example, $M_{g0} < M_{gA} = 12.28$, and therefore the primary
subshock solution appears when the low-velocity root
$u_- > u_s$, which occurs when $M_{c0} > M_{cA} = 4.25$.
The same function is plotted on a smaller scale in
({\it b}), where we see that two new subshock roots for $u_-$
appear when $M_{c0} > 41.65$.

\bigskip

Fig. 6. -- Function $h(u_-)$ (eq.~[4.8]) is plotted for the parameters
({\it a}) $M_{g0} = 13$, $\gamma_g = 5/3$, $\gamma_c = 4/3$.
The value of $M_{c0}$
is indicated for each curve. In this example,
$M_{g0} > M_{gA} = 12.28$, and therefore the primary
subshock solution never appears. Hence smooth flow is
possible for all values of $M_{c0}$. The same function is plotted
on a smaller scale in ({\it b}), where we see that two new subshock
roots for $u_-$ appear when $M_{c0} > 116$.

\bigskip

Fig. 7. -- Critical Mach numbers $M_{gB}$ and $M_{cB}$ for the onset of
multiple solutions (eqs.~[4.11] and [4.12]) are plotted as parametric
functions of the pre-subshock velocity $u_-$ in the $(M_{g0},M_{c0})$
parameter space for ({\it a}) $\gamma_g = 5/3$ and the indicated values
of $\gamma_c\,$; ({\it b}) $\gamma_c = 4/3$ and the indicated values of
$\gamma_g$. The interior of each wedge is the multiple-solution region
for the associated parameters.

\bigskip

Fig. 8. -- Critical upstream Mach numbers for the occurrence of
multiple solutions (eqs.~[4.11] and [4.12]; {\it solid line}) and
for smooth flow (eq.~[3.4]; {\it dashed line}) are plotted together
in the $(M_{g0},M_{c0})$ parameter space for the case $\gamma_g = 5/3$
and $\gamma_c = 4/3$. The minimum upstream cosmic-ray Mach number
required for decelerating flow is also shown (eq.~[2.32];
{\it dotted line}). There are four distinct domains in the
parameter space as discussed in the text.

\bigskip

Fig. 9. -- Function $h(u_-)$ (eq.~[4.8]) is plotted for the parameters
({\it a}) $M_{g0} = 6.5$, $\gamma_g = 5/3$, $\gamma_c = 4/3$ along
the segment $RP$ in Fig.~8. The values of the upstream cosmic-ray
Mach number are
$M_{c0} = 45$ ({\it solid line}),
$M_{c0} = 26$ ({\it dashed line}),
$M_{c0} = 15$ ({\it dotted line}). When $M_{c0} > M_{cB} = 26$,
there are three distinct subshock solutions available. In panel
({\it b}) the same function is plotted on a smaller scale.

\bigskip

Fig. 10. -- Function $h(u_-)$ (eq.~[4.8]) is plotted for the parameters
({\it a}) $M_{c0} = 45$, $\gamma_g = 5/3$, $\gamma_c = 4/3$ along the segment
$QP$ in Fig.~8. The values of the upstream gas Mach number are
$M_{g0} = 6.5$ ({\it solid line}),
$M_{g0} = 5.83$ ({\it dashed line}),
$M_{g0} = 4.8$ ({\it dotted line}). When $M_{g0} > M_{gB} = 5.83$,
there are three distinct subshock solutions available. In panel
({\it b}) the same function is plotted on a smaller scale.

\bigskip

Fig. 11. -- Critical upstream Mach numbers for the occurrence of
multiple solutions (eqs.~[4.11] and [4.12]; {\it solid line}) and
for smooth flow (eq.~[3.4]; {\it dashed line}) are plotted together
in the $(M_{g0},M_{c0})$ parameter space for
({\it a}) $\gamma_g = 5/3$, $\gamma_c = 4/3$;
({\it b}) $\gamma_g = 5/3$, $\gamma_c = 1.35$;
({\it c}) $\gamma_g = 1.6$, $\gamma_c = 4/3$;
({\it d}) $\gamma_g = 1.6$, $\gamma_c = 1.35$.
Also indicated is the minimum value of $M_{c0}$ required for
decelerating flow (eq.~[2.32]; {\it dotted line}).

\bigskip

Fig. 12. -- Our analytical results for the critical curves generated
using equations~(3.4), (4.11), and (4.12) are combined with equations~(1.5)
to create corresponding curves in the alternative parameter spaces
$(M_0,Q_0)$ and $(M_{g0},Q_0)$ employed by Bulanov \& Sokolov (1984)
and Ko, Chan, \& Webb (1997), respectively. Panel ({\it a}), with
$\gamma_g = 5/3$ and $\gamma_c = 4/3$, is identical
to Fig.~4 of Bulanov \& Sokolov (1984).
Panel ({\it b}), with $\gamma_g = 2$ and $\gamma_c = 4/3$, is
identical to Fig.~1({\it a}) of Ko, Chan, \& Webb (1997).
Note that these authors generated their curves using root-finding
procedures. The interpretation of the line styles is the same as in
Fig.~11.

\bigskip

Fig. 13. -- Numerical solutions for ({\it a}) $u$, ({\it b}) $M_g$,
({\it c}) ${\cal P}_g$, and ({\it d}) ${\cal P}_c$ are plotted as functions of
$\tau$ (see eq.~[5.3]). The solutions were obtained by integrating the
dynamical eq.~[5.1] with $M_{g0} = 4$ and $M_{c0} = 4$, which
corresponds to Domain I in Fig.~8. In this
case one discontinuous solution is possible, and smooth flow
is impossible.

\bigskip

Fig. 14. -- Same as Fig.~13, except $M_{g0} = 6$ and $M_{c0} = 60$,
which corresponds to Domain II in Fig.~8.
In this case three distinct discontinuous solutions are possible,
and smooth flow is impossible. The values of the pre-subshock gas
Mach number are
$M_{g-} = 2.54$ ({\it solid line}),
$M_{g-} = 3.84$ ({\it dashed line}),
$M_{g-} = 5.94$ ({\it dotted line}).

\bigskip

Fig. 15. -- Same as Fig.~13, except $M_{g0} = 12.5$ and $M_{c0} = 200$,
which corresponds to Domain III in Fig.~8.
In this case two distinct discontinuous solutions are possible
in addition to one globally smooth solution ({\it solid line}).
The values of the pre-subshock gas Mach number are
$M_{g-} = 11.9$ ({\it dashed line}),
$M_{g-} = 12.4$ ({\it dotted line}).

\bigskip

Fig. 16. -- Same as Fig.~13, except $M_{g0} = 14$ and $M_{c0} = 20$,
which corresponds to Domain IV in Fig.~8.
In this case one globally smooth solution solution is possible,
and discontinuous flow is impossible.


\begin{thebibliography}{}

\bibitem{} Achterberg, A. 1987, Astron. Ap., 174, 329

\bibitem{} Achterberg, A., Blandford, R., \& Periwal, V. 1984, Astron. Ap.,
132, 97

\bibitem{} Axford, W. I., Leer, E., \& McKenzie, J. F. 1982, Astron. Ap.,
111, 317

\bibitem{} Axford, W. I., Leer, E., \& Skadron, G. 1977, 15th Int. Conf.
Cosmic Rays (Plovdiv), 11, 132

\bibitem{} Becker, P. A. 1998, ApJ, 498, 790

\bibitem{} Bell, A. R., 1978a, MNRAS, 182, 147

\bibitem{} Bell, A. R., 1978b, MNRAS, 182, 443

\bibitem{} Blandford, R. D., \& Ostriker, J. P. 1978, ApJ, 221, L29

\bibitem{} Bulanov, S. V., \& Sokolov, I. V. 1984, Sov. Astron.,
28, 515

\bibitem{} Donohue, D. J., Zank, G. P., \& Webb, G. M. 1994, ApJ, 424, 263

\bibitem{} Drury, L. O'C., \& Falle, S. A. E. G. 1986, MNRAS, 223, 353

\bibitem{} Drury, L. O'C., \& V\"olk, H. J. 1981, ApJ, 248, 344 (DV)

\bibitem{} Duffy, P., Drury, L. O'C., \& V\"olk, H. 1994, Astron. Ap.,
291, 613

\bibitem{} Frank, A., Jones, T. W., \& Ryu, D. 1995, ApJ, 441, 629

\bibitem{} Gleeson, L. J., \& Axford, W. I. 1967, ApJ, 149, L115

\bibitem{} Heavens, A. F. 1984a, MNRAS, 207, 1P

\bibitem{} Heavens, A. F. 1984b, MNRAS, 210, 813

\bibitem{} Jones, F. C., \& Ellison, D. C. 1991, Space. Sci. Rev., 58, 259

\bibitem{} Kang, H., \& Jones, T. W. 1990, ApJ, 353, 149

\bibitem{} Kang, H., Jones, T. W., \& Ryu, D. 1992, ApJ, 385, 193

\bibitem{} Kirk, J. G., \& Webb, G. M. 1988, ApJ, 331, 336

\bibitem{} Ko, C.-M. 1995a, MNRAS, 275, 1211

\bibitem{} Ko, C.-M. 1995b, Adv. Space Res., 15, 149

\bibitem{} Ko, C.-M., Chan, K.-W., \& Webb, G. M. 1997, J. Plasma Physics,
57, 677

\bibitem{} Krymskii, G. F. 1977, Sov. Phys. Dokl., 22, 327

\bibitem{} Landau, L. D., \& Lifshitz, E. M. 1975, Fluid Mechanics
(NY: Pergamon)

\bibitem{} Malkov, M. A. 1997a, ApJ, 485, 638

\bibitem{} Malkov, M. A. 1997b, ApJ, 491, 584

\bibitem{} Malkov, M. A., \& V\"olk, H. J. 1996, ApJ, 473, 347

\bibitem{} McKenzie, J. F., \& V\"olk, H. J. 1982,
Astron. Ap., 116, 191

\bibitem{} Mond, M., \& Drury, L. O'C. 1998, Astron. Ap.,
332, 385

\bibitem{} Reif, F. 1965, Fundamentals of Statistical and Thermal
Physics (NY: McGraw-Hill)

\bibitem{} Ryu, D., Kang, H., \& Jones, T. W. 1993, ApJ, 405, 199

\bibitem{} Skilling, J. 1971, ApJ, 170, 265

\bibitem{} Skilling, J. 1975, MNRAS, 172, 557

\bibitem{} V\"olk, H. J., Drury, L. O'C., \& McKenzie, J. F. 1984,
Astron. Ap., 130, 19

\bibitem{} Zank, G. P., Axford, W. I., \& McKenzie, J. F. 1990, Astron. Ap.,
233, 275

\bibitem{} Zank, G. P., Webb, G. M., \& Donohue, D. J. 1993, ApJ, 406, 67

\end{thebibliography}
\end{document}